\documentclass[12pt,preprint]{aastex}
\newcommand{\kms}{km s$^{-1}\;$}

\begin{document}
\title{Bright Variable Stars in NGC 6819 - An Open Cluster in the {\it Kepler} Field}
\author{Antonio Talamantes, Eric L. Sandquist}
\affil{San Diego State University, Department of Astronomy, San Diego,
CA 92182}
\email{atalaman@pleiades.sdsu.edu,erics@mintaka.sdsu.edu}

\author{James L. Clem} 
\affil{Department of Physics \& Astronomy, Louisiana State University,
Baton Rouge, LA 70803-4001} 
\email{jclem@phys.lsu.edu}

\author{Russell M. Robb, David D. Balam}
\affil{Department of Physics and Astronomy, University of Victoria, Victoria, 
BC, Canada, V8W 3P6}
\email{robb@uvic.ca,cosmos@uvic.ca}

\author{Matthew Shetrone}
\affil{University of Texas, McDonald Observatory, HC75 Box 1337-L 
    Fort Davis, TX, 79734}
\email{shetrone@astro.as.utexas.edu}

\begin{abstract}
We describe a variability study of the moderately old open cluster NGC
6819. We have detected 4 new detached eclipsing binaries near the cluster
turnoff (one of which may be in a triple system). Several of these systems
should be able to provide mass and radius information, and can therefore
constrain the age of the cluster. We have also newly detected one possible
detached binary member about 3.5 magnitudes below the turnoff. One EW-type
binary (probably not a cluster member) shows unusually strong
night-to-night light curve variations in sets of observations separated by 8
years.

According to the best current information, the three brightest
variables we detected (2 of them new) are cluster members, making them
blue stragglers. One is a $\delta$ Scu pulsating variable, one is a
close but detached binary, and the third contains a detached short
period binary that shows total eclipses. In each case, however, there
is evidence hinting that the system may have been produced through the
interaction of more than two stars.
\end{abstract}


\section{Introduction}

Star clusters have long been used as laboratories for testing our
understanding of the physics that goes into computing models of
stars. Most commonly, the brightnesses and colors of the stellar
population have been compared with isochrones to check our
understanding of how stars of nearly the same age and chemical
composition evolve.  Unfortunately, observational and theoretical
issues prevent highly accurate measurements of desirable quantities
like age. For example, convection and color-$T_{\rm eff}$ relations
present notorious problems in modeling the light emitted by stellar
populations.  Degeneracies in distance, interstellar reddening,
chemical composition greatly complicate efforts to measure ages as well
\citep[e.g.]{south}.

The different methods of deriving information about the stars in a population
can magnify each other. The analysis of eclipsing binary stars can
provide accurate mass and radius information on a subset of the stars,
and when used in conjunction with colors and brightnesses, this can
help to ``pin down'' theoretical isochrones at a number of
locations. This immensely increases the constricting power of the
observations. Asteroseismology can provide accurate mean densities for
individual stars and holds promise of revealing details of their
internal structure. But again, when these observation can be used with
color-magnitude diagram information the power of the tests are
magnified.

Star clusters in the field of the {\it Kepler} satellite promise to
become important laboratories for stellar astronomy because of the
range of observations that can be brought to bear to test theoretical
models. Asteroseismological oscillations have already been detected in
giant stars in the cluster NGC 6819 \citep{stello}, and efforts are on
to detect them among fainter stars. This paper is the start of our
attempt to identify detached eclipsing binary stars in this same
cluster. {\it Kepler} will provide light curves of unprecedented
accuracy that will greatly improve the precision of radius
measurements for stars in the cluster, but to make full use of the
binary systems, we also need temperature information that will come
from observations in narrower wavelength bands. In turn, masses for
turnoff stars will help directly constrain the low-mass end of the
white dwarf initial-final mass relation \citep[e.g.]{kalirai,will},
which in turn impacts chemical enrichment and white dwarf cooling
studies.

NGC 6819 is a moderately old cluster ($\sim 2.5$ Gyr; \citealt{ros,kalirai})
with near-solar or slightly super-solar metallicity ([Fe/H]$=+0.09 \pm 0.03$;
\citealt{brag}).  Variability studies have been presented previously for NGC
6819 by \citet{ks}, \citet{street1}, and \citet{street2}. The wider surveys of
\citet{hart} and \citet{pig} also covered large portions of the {\it Kepler}
field as part of the Hungarian-made Automated Telescope network (HATnet)
project and the All-Sky Automated Survey (ASAS), respectively. Both the HATnet
and ASAS surveys had low spatial resolution ($14\arcsec$ and $15\arcsec$ per
pixel) and were relatively shallow, so there is no overlap with the variables
we discuss below. \citet{ks} were searching for short period contact binaries
in the cluster, and their observing program only involved two nights. Even so
they identified three possible variables, two of which we have confirmed. (The
third, which corresponds to star A81\footnotetext{Identification numbers in
  this paper that start with ``A'' come from the photometric study by
  \citet{auner}.}, was not seen to vary in our datasets.)  In a series of
  papers, Street et al. identified a large number of variables in and near the
  cluster as a byproduct of a planet search. Their exposures resulted in
  saturation for stars with $V \la 16.5$, which is fainter than the cluster
  turnoff. There was therefore a distinct possibility that undetected detached
  eclipsing variables were present in the cluster.


While characterization of the eclipsing binaries is our primary goal,
accurate and precise stellar photometry needs to be of complementary quality.
\citet{auner} presented the first wide field photographic photometry
of NGC 6819, while the primary CCD photometry has been published by
\citet{ros} and \citet{kalirai}.  The photometry in this study covers
a slightly larger field and goes slightly fainter than that of
\citet{ros}. By contrast, \citeauthor{kalirai}'s study covered a much
wider field and went deep enough to identify the cluster's white dwarf
sequence.

\section{Observations and Data Reduction}

NGC 6819 was observed using two different telescopes. Some of us (J.L.C.,
R.M.R, D.D.B) observed the cluster for 10 consecutive nights using the 1.8
meter Plaskett Telescope at Dominion Astrophysical Observatory (hereafter,
DAO) between July 5-15, 2001 (although weather compromised most of the night
of July 7). These images were taken using a Cousins $R$ filter (hereafter
$R_C$) and lasted for a total of 180 s each. The field of view, spanning a
total area of $9\farcm2\times 9\farcm2$. A 1024$\times$1024 CCD was used, and
pixels were binned $2\times 2$, for a final $1\farcs1$ per pixel. 16 nights of
data were taken at the Mount Laguna Observatory (hereafter, MLO) 1 m telescope
(by A.T. and E.L.S.) covering an area approximately $13\farcm5 \times
13\farcm5$
, for a scale of about $0\farcs4$ per pixel.
and are listed in Table \ref{phottab}. For eleven of the nights the exposures
were taken using a $V$ filter and typically lasted 90 s each.  The remaining
six nights of observations used a Kron $R$ filter (hereafter $R_K$, which has
a passband similar to the combination of Cousins $R$ and $I$) and again each
exposure was typically 90 s. Two additional nights of observations were taken
at MLO for the purpose of calibrating the photometry to the standard
system. For both the DAO and MLO datasets, the seeing typically produced a
FWHM of stellar images around $2-2.5$ arcsec. Image reduction (overscan
subtraction, bias subtraction, and flat field correction) was done using
standard tools in IRAF\footnote{IRAF (Image Reduction and Analysis Facility)
  is distributed by the National Optical Astronomy Observatories, which are
  operated by the Association of Universities for Research in Astronomy, Inc.,
  under contract with the National Science Foundation.}.

Differential photometry was undertaken using the image subtraction package
ISIS \citep{isis}. ISIS is commonly used in variability studies today, thanks
to its ability to detect variability at near the Poisson noise limit.  Our
image sets in $V$, $R_C$, and $R_K$ filters were interpolated to a common
image coordinate system (separately for the DAO and MLO datasets), and then
processed separately by filter. This involved producing a reference image from
approximately 20 images with the best seeing; subtracting each interpolated
image from the reference image once the reference was transformed to the
seeing of the interpolated image (and a constant flux scaling is enforced);
stacking the subtracted images to identify stars whose detected flux had
changed; and taking photometric measurements of the subtracted image. While
ISIS has the capability of breaking images into subimages and analyzing the
point-spread function separately on each subimage, we found that using whole
images produced better results for our datasets.

Several routines of the ISIS code do not function optimally,
however. The ISIS code uses a bicubic spline to interpolate each image
to the coordinate system set by an astrometric reference
image. Bicubic splines are subject to spurious oscillations near sharp
features though, so we replaced the bicubic spline with an Akima
spline. This reduced noise that had been introduced by saturated stars
and diffraction spikes, but also improved the interpolation of the
images of bright stars on images with good seeing.

The output of ISIS is a difference flux measured on the subtracted
images.  To convert these difference fluxes into magnitudes, the
star's flux must be measured on the reference frame. For crowded star
fields, this should ideally be done using a technique like profile
fitting that is capable of accurate photometry in the limit of
strongly overlapping star images. An important issue is that a flux
derived in that way is not guaranteed to be on the same scale as the
ISIS difference flux \citep{hart}, and if it is not, this will
directly affect the amplitude of derived light curves. 
For our datasets, we found that a reference flux derived from ISIS' own
photometry routine was a reliable choice. We are principally interested in
ensuring that our {\it differential} photometry is accurate because this
affects the reliability of stellar sizes computed from light curve analysis.
To this end, we subtracted our aperture photometry results from ISIS
photometry point-by-point for a selection of eclipsing binary stars that were
not affected by blending issues. When done, it was impossible to distinguish
between points in or out of eclipses.

ISIS' algorithm is a modified aperture photometry routine that
employs the reference image point-spread function (PSF) for weighting
purposes. The reference PSF is first transformed to the seeing of the
image under consideration. Only the portion within an aperture of {\tt
  radphot} pixels is used, but it is normalized to a larger aperture
of {\tt rad\_aper} pixels. The pixel values in the subtracted image
are weighted by this transformed PSF. For the MLO datasets, we used
{\tt radphot} $= 4$ pix and {\tt rad\_aper} = 10 pixels. For the DAO
images (which had pixels with larger area on the sky), we used {\tt
  radphot} $= 2$ pix and {\tt rad\_aper} = 7. The photometry routine
requires coordinate lists for the the stars to be measured --- both
centroid and pixel with maximum counts. As written, ISIS uses the
pixel with maximum counts as the center of the apertures. It should be
noted that the ISIS starts its pixel coordinate system at $x=y=0$, unlike
many astronomical software packages, and this should be accounted for
if producing a coordinate list outside of ISIS' routines.

\begin{figure} 
\epsscale{0.7}
\plotone{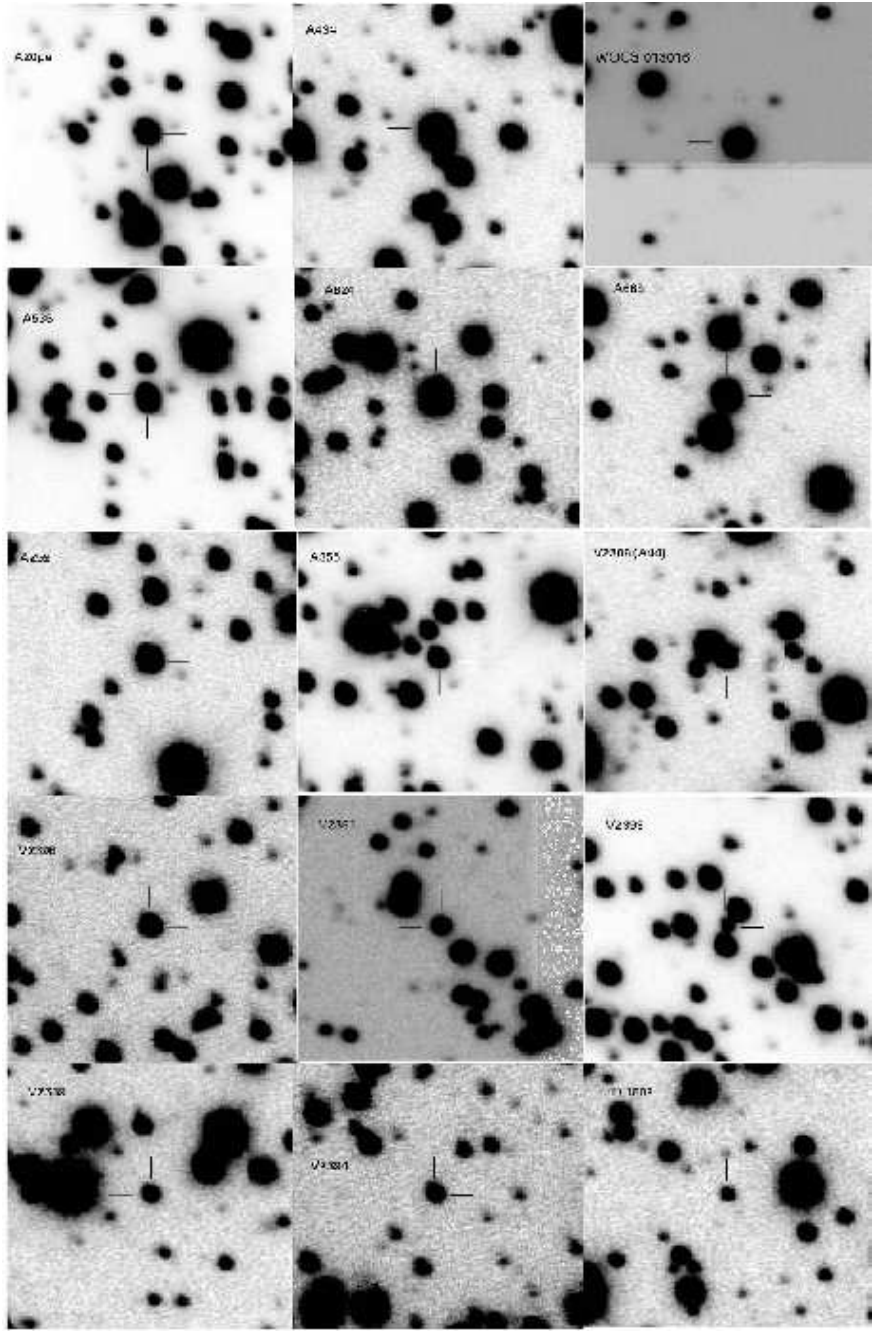}
\caption{Finder charts in $V$ band for the variables in NGC 6819 described in
  this paper. Individual squares are approximately $1\arcmin$ across, and are
  oriented so north is at the top and east is to the left.\label{pic}}
\end{figure}

Finder charts for the detected variables are shown in Fig. \ref{pic}.

\subsection{Photometric Calibration}

We took calibration images under photometric conditions on the night of 25
October 2008. We observed the standard fields PG0231+051, SA 92, SA95, SA 98,
and NGC 6940, and used standard values taken from \citet[][retrieved August
2009]{stet}. The standard fields were observed between 3 and 10 times per
filter in $BVI$ at airmasses that ranged from 1.034 to 1.998. All together
this resulted in more than 4000 standard star observations per filter covering
a color range $-0.5 \la (B-I) \la 5$.

\begin{figure}
\plotone{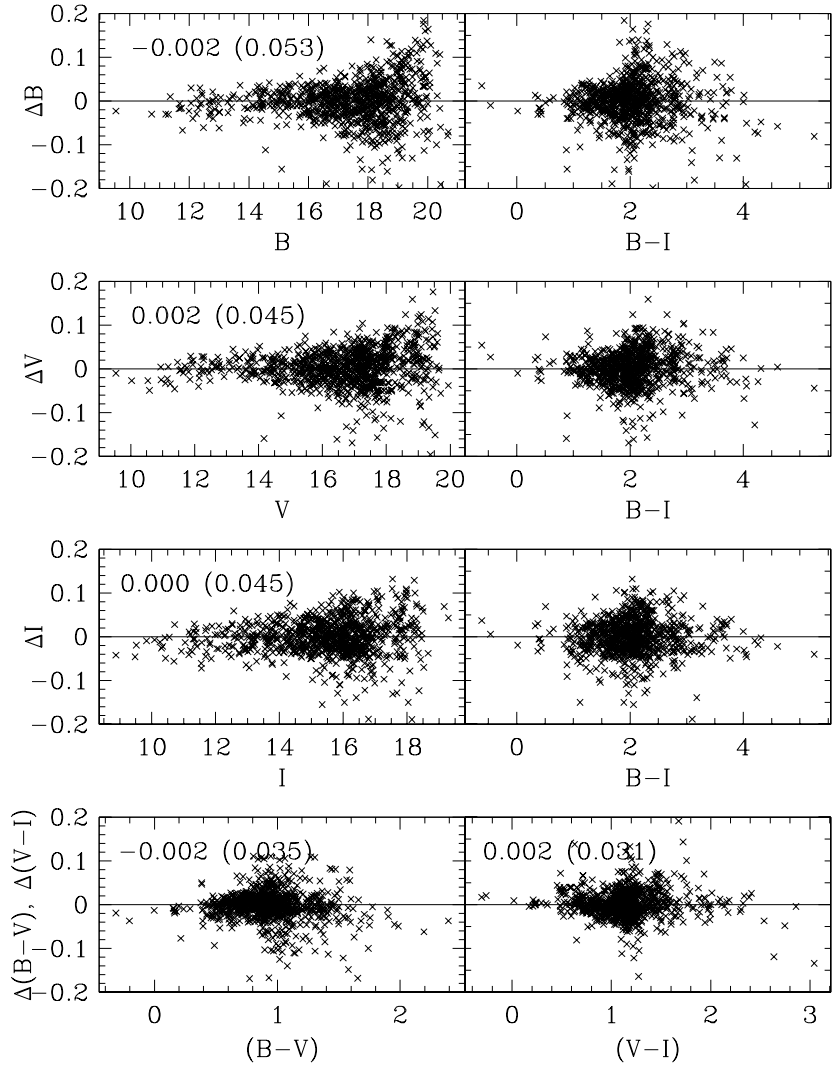}
\caption{Residuals between our calibrated photometry of photometric standards
  and the standard magnitudes of \citealt{stet} (in the sense of this study
  minus Stetson's).\label{stetcomp}}
\end{figure}

We derived aperture photometry from all frames using DAOPHOT, and made curve
of growth corrections using the program DAOGROW.
The observations were transformed to the standard system using the
following equations:
\[ b = B + a_0 + a_1 (B-I) + a_2 X \]
\[ v = V + b_0 + b_1 (B-I) + b_2 X \]
\[ i = I + c_0 + c_1 (B-I) + c_2 X \]
where $b, v,$ and $i$ are instrumental magnitudes, $B, V$, and $I$ are
standard-system magnitudes, $X$ is airmass, and $a_i, b_i$, and $c_i$ are
coefficients determined from least-squares fits. For the color coefficients,
we found $a_1 = -0.0208 \pm0.0018$, $b_1 = 0.0367 \pm 0.0007$, and $c_1 =
-0.0066 \pm 0.0006$. The extinction coefficients were $a_2 = 0.233 \pm 0.004$,
$b_2 = 0.129 \pm 0.005$, and $c_2 = 0.064 \pm 0.007$. Fig. \ref{stetcomp}
shows the residuals of the comparison of our standardized observations and the
Stetson values.

Clusters to be calibrated were observed with a range of exposures times on the
same night. For NGC 6819, there were 8 observations in $B$ ($1\times30$ s,
$2\times60$ s, $2\times120$ s, $2\times180$ s, and $1\times300$ s), 9
observations in $V$ ($2\times10$ s, $2\times60$ s, $4\times120$ s, and
$1\times300$ s), and 11 observations in $I$ ($4\times10$ s, $1\times60$ s,
$5\times120$ s, and $1\times300$ s).

\begin{figure}
\plotone{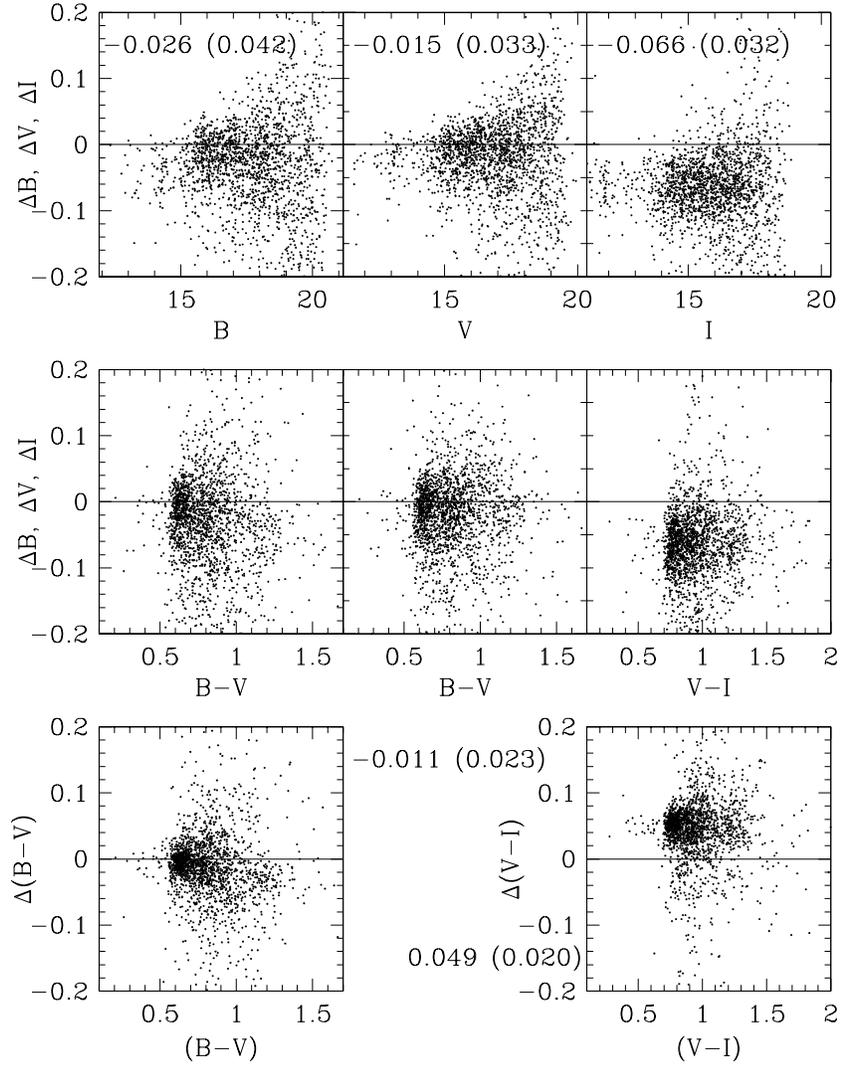}
\caption{Residuals between our photometry of NGC 6819 and that of
  \citet{ros}, in the sense of this study minus theirs. The quoted numbers are
  the median residual and the semi-interquartile range.\label{rvcomp}}
\end{figure}

\begin{figure}
\epsscale{0.9}
\plotone{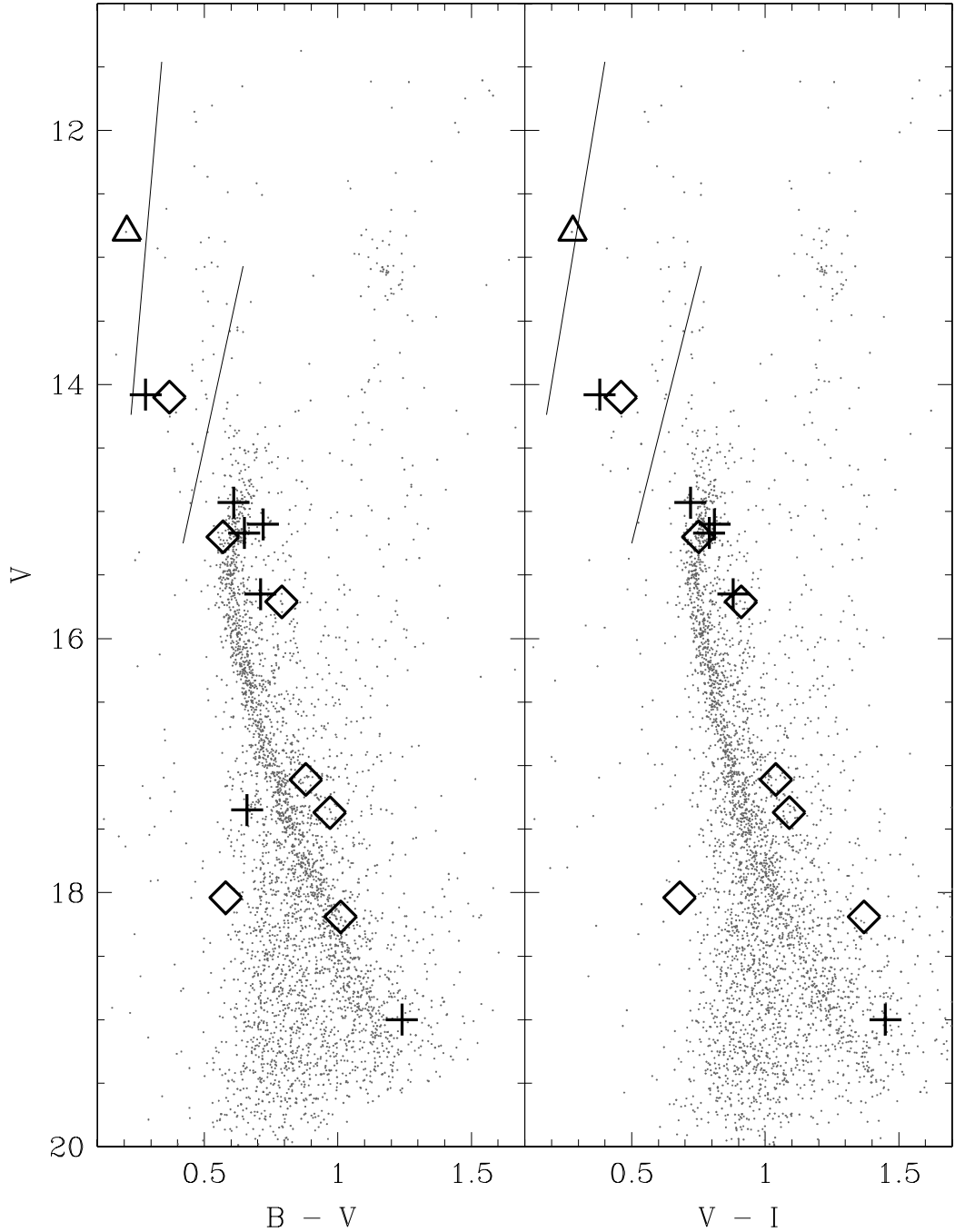}
\caption{Color-magnitude diagram for NGC 6819 with variables identified.
  Detached eclipsing binaries are shown with crosses, contact and near contact
  binaries are shown with diamonds, and pulsating variables are shown with
  triangles. Solid lines are the theoretical instability strip from
  \citet{pam}.\label{cmd}}
\end{figure}

Figure \ref{rvcomp} shows a comparison of our photometry with that of
\citet{ros}, as downloaded from the WEBDA database. It should be noted that
\citeauthor{ros} did not present the details of the $I_C$ observations or
reduction in their article, Overall there is no sign of color-dependent
residuals and there is very good agreement in zeropoints, with the most
noticeable difference being a relatively large zeropoint difference in $I_C$.
The color-magnitude diagram for our field is shown in Fig. \ref{cmd}.

\subsection{Spectra}

Our spectra were obtained at the Hobby-Eberly Telescope (HET) with the
High Resolution Spectrograph (HRS, \citealt{tull}) as part of normal
queue scheduled observing \citep{shetet}.  HRS was configured
to HRS\_30k\_central\_316g5936\_2as\_0sky\_IS0\_GC0\_2x3 or
HRS\_30k\_central\_600g5822\_2as\_2sky\_IS0\_GC0\_2x3 to achieve
R=30,000 spectra covering 4100\AA\ to 7800\AA\ or 4825\AA\ to
6750\AA\, respectively.  Exposure times ranged from 600 seconds to
1200 seconds.  The signal-to-noise was typically 25 per resolution
element at 5800 \AA.

The spectra were reduced with IRAF ECHELLE scripts.  The standard IRAF
scripts for overscan removal, bias subtraction, flat fielding and
scattered light removal were employed.  For the HRS flat field we
masked out the Li I, H I and Na D regions because the HET HRS flat
field lamp suffered from faint emission lines.  The spectra were
combined into a single long spectrum for the blue and red chips.
Radial velocities were determined from cross-correlation using the
IRAF task fxcor using a solar spectrum as template.  The
heliocentric correction was made using the IRAF task rvcorrect.

\section{Results}


\subsection{Pulsating Star A20pe}

This star\footnote{This is photoelectric standard 20 in \citet{auner},
  not part of his photographic photometry list. The identifications of
  this star seem to be confused in the WEBDA database. This is star 36
  in the photometry of \citet{ros}, but that star is incorrectly
  identified with star 168 in the \citet{pm} proper motion study.} has
identification number 107 in the proper motion study of \citet{pm},
which gives it a 90\% probability of membership. It is as bright in
$V$ as stars in the red clump for the cluster, but is bluer than the
cluster turnoff. A20pe is therefore a likely cluster blue straggler.

The star has low-amplitude short period variation that identifies it as a
$\delta$ Scu pulsating variable. As such, it should be within the instability
strip. We transformed the theoretical instability strip boundaries of
\citet{pam} to the observational plane using color-$T_{\rm eff}$ relations
from \citet{vc03}, and $E(B-V) = 0.10$ and $(m-M)_V = 12.30$
\citep{kalirai}. Based on this, the star appears to be too blue and/or
bright. It is possible that the star could have a companion that is putting it
out of the strip --- a direct analog to this idea exists in the blue straggler
EX Cnc in M67, which is an eccentric spectroscopic binary of 4.2 d period
containing a $\delta$ Scuti variable \citep{ml}. However, in this case, the
companion star would have to be {\it hotter} and more massive than the
$\delta$ Scu star, implying that there would be two stars in a binary that
would be identified as blue stragglers on their own. Radial velocities would
be helpful in verifying the star's membership in the cluster, and if a member,
helping to identify its formation mechanism.


\subsection{Contact or Near-Contact Binaries}

\begin{figure}
\plotone{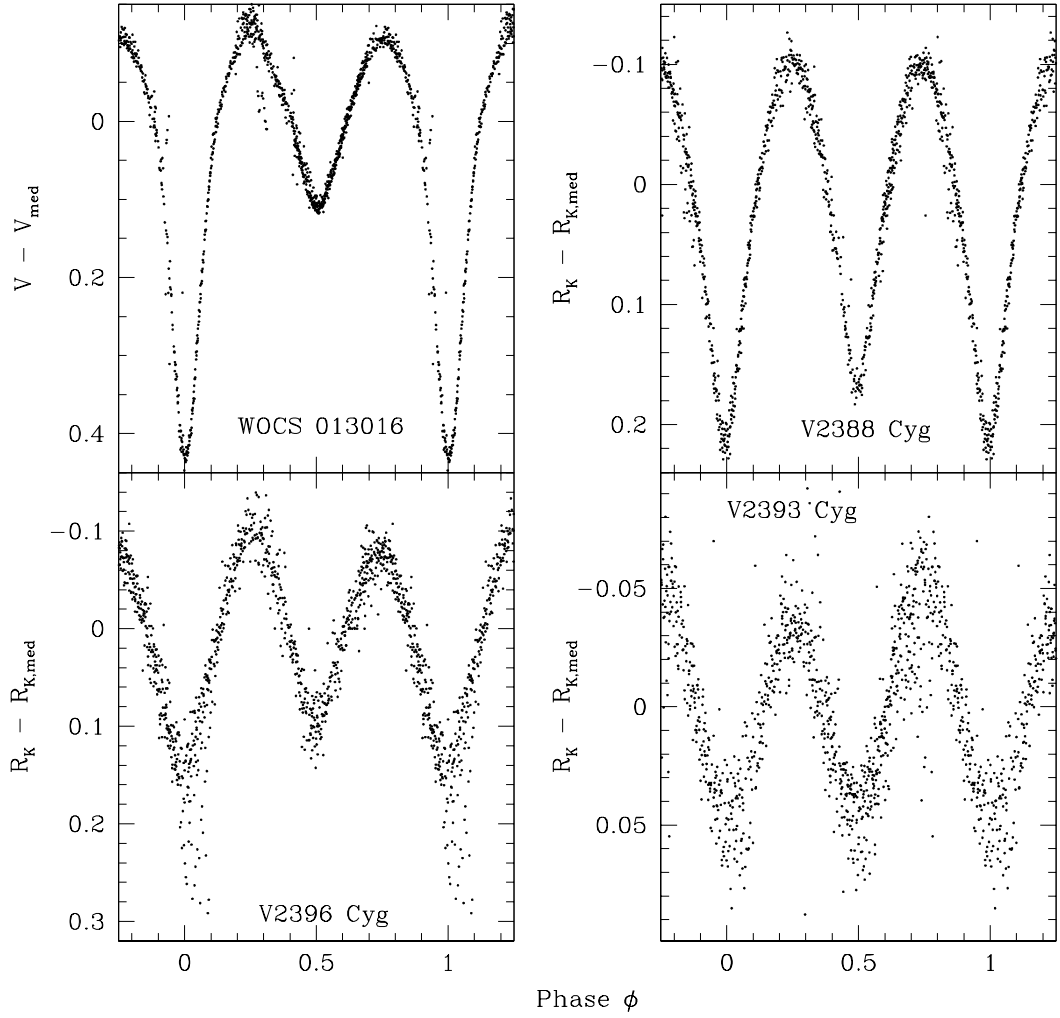}
\caption{Phased light curves for contact and near-contact binaries.\label{ews}}
\end{figure}

Phased light curves for some of the short-period eclipsing binaries
are shown in Fig. \ref{ews}.  Where possible, the photometry presented
in Table \ref{vartab} has been corrected to system maximum
values. Most of these systems were previously identified by
\citet{street1}, and so we focus here on new systems and unusual
properties of the known systems that have not been discussed before.

Before discussing those, we attempted to estimate likelihood of
membership using the color-luminosity relationship for W UMa variable
stars from \citet{ruc97}:
\[ M_V = -4.44 \log P + 3.02 (B-V)_0 + 0.12 \] 
and then comparing the implied distance modulus to the one derived from
isochrone fitting. $(B-V)_0$ is the dereddened color of the system at maximum
brightness. \citet{kalirai} derived a distance modulus $(m-M)_V = 12.30 \pm
0.12$ and assumed a reddening $E(B-V) = 0.10$. Given the uncertainty in the
distance modulus and the observed scatter of 0.22 mag in the \citet{ruc97}
relationship, V2396 Cyg [$(m-M)_V = 12.26$], V2388 Cyg (12.53) are
probable cluster members, while A355 (11.47) and V2389 Cyg (14.39) are not.
We discuss the special case WOCS 013016 immediately below.

{\bf WOCS 013016:} This is a new detection because it was too bright
for the variability surveys by \citet{street1,street2} and was outside
the field studied by \citet{ks}. The $V$ photometry presented in Table
\ref{vartab} has been corrected to maximum system brightness, but the
colors are somewhat less accurate --- our calibration observations
occurred near, but not at maximum light.

The system has identification number 46 in the proper motion study of
\citet{pm}, which gives it a 80\% probability of membership.  If we apply the
\citet{ruc97} period-luminosity-color (PLC) relationship here, we calculate
$(m-M)_V = 12.38$, in very good agreement with the cluster distance modulus.
However, WOCS 013016 has a period that is significantly longer than the
probable cluster EW systems that we detected, and its light curve more closely
resembles those of EB systems.  \citet{ruc97} discuss the application of their
period-luminosity-color (PLC) relationship to EW systems that appear to be in
poor thermal contact (in other words, having eclipses of significantly
different depths, implying significant differences in the surface temperatures
of the components). They indicate that the PLC relationship should produce
luminosity estimates that are too high.
As long as the biases are not too large ($\la 0.3$ mag), the distance modulus
for WOCS 013016 supports its membership in the cluster.

If it is a member, then it should be identified as a blue straggler
that is probably in the process of coalescing. To be as blue as it is
currently, there must have been an earlier merger or significant mass
transfer to have produced the more massive component of the binary.

\begin{figure} 
\plotone{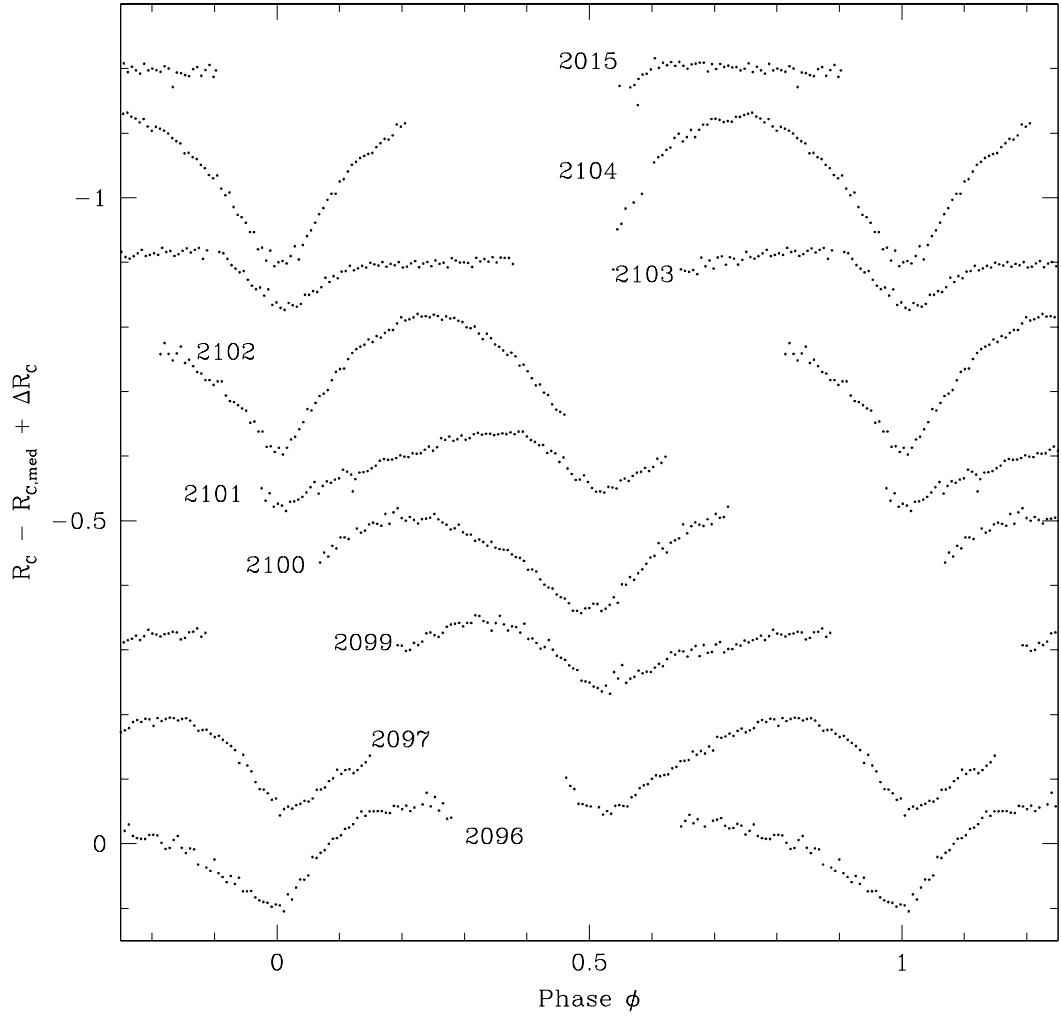}
\caption{Phased light curves from individual nights of observations in
  $R_C$. Numerical labels corresponds to the heliocentric Julian date at the
  start of observations minus 2450000. Each night was shifted brightward in
  the plot by 0.15 mag from the previous night plotted.\label{a355}}
\end{figure}

{\bf A355 (WOCS 052004):} This variable is brighter than the saturation limit
for images taken by \citet{street1}, but was detected earlier by
\citet{ks}.\footnote{This star is labeled as ``V2'' in the finder chart of
  \citet{ks}, but their tabulated photometry and the position of V2 in their
  CMD is not consistent with our photometry. We believe that ``V1'' in their
  table and CMD corresponds to this variable.}  This binary system is probably
in contact, but it shows {\it unusually severe} distortions to its light
curve, as seen in Fig. \ref{a355}. Similar behavior was seen also in $V$ and
$R_K$ filter bands with a different telescope/camera set-up years later, so
this behavior is very likely to be real and ongoing. For example, in several
cases in the figure (and on the nights HJD 2454697 to 2454699), the
photometric maximum is found as much as 0.1 cycle away from the expected
position, and it shifts significantly from night to night. In other cases
(HJDs 2452103 and 2452105 in the figure, as well as HJDs 2454624 and 2454669),
one of the photometric maximums was almost completely truncated, to the point
of making the system superficially resemble a detached eclipsing binary of
nearly spherical stars. Before and after night 2452103, the system showed some
of its deepest eclipses and showed variations identifying it as an EW system
with gravitationally distorted stars of nearly equal temperature.  In nearly
all cases where we can compare phases 0.5 apart on the same night, the light
curves imply the system is nearly point symmetric about the center of the
orbits, but not symmetric about the line connecting the star centers.  This
type of variation is difficult to understand in terms of spots in part because
of the symmetry and in part because of the short timescales.

The distortions of the light curve made it difficult to derive a period, but
the value given in Table \ref{vartab} does a good job of align the eclipses in
phase. While we attempted to correct the system photometry to maximum light,
we do not have light curve information during the calibration observations, so
that we are not full able to judge what kind of variations might have been
occuring. While the distortions of the light curve are interesting, because
the object does not appear to be a cluster member, we do not discuss it
further here.

\begin{figure}
\plotone{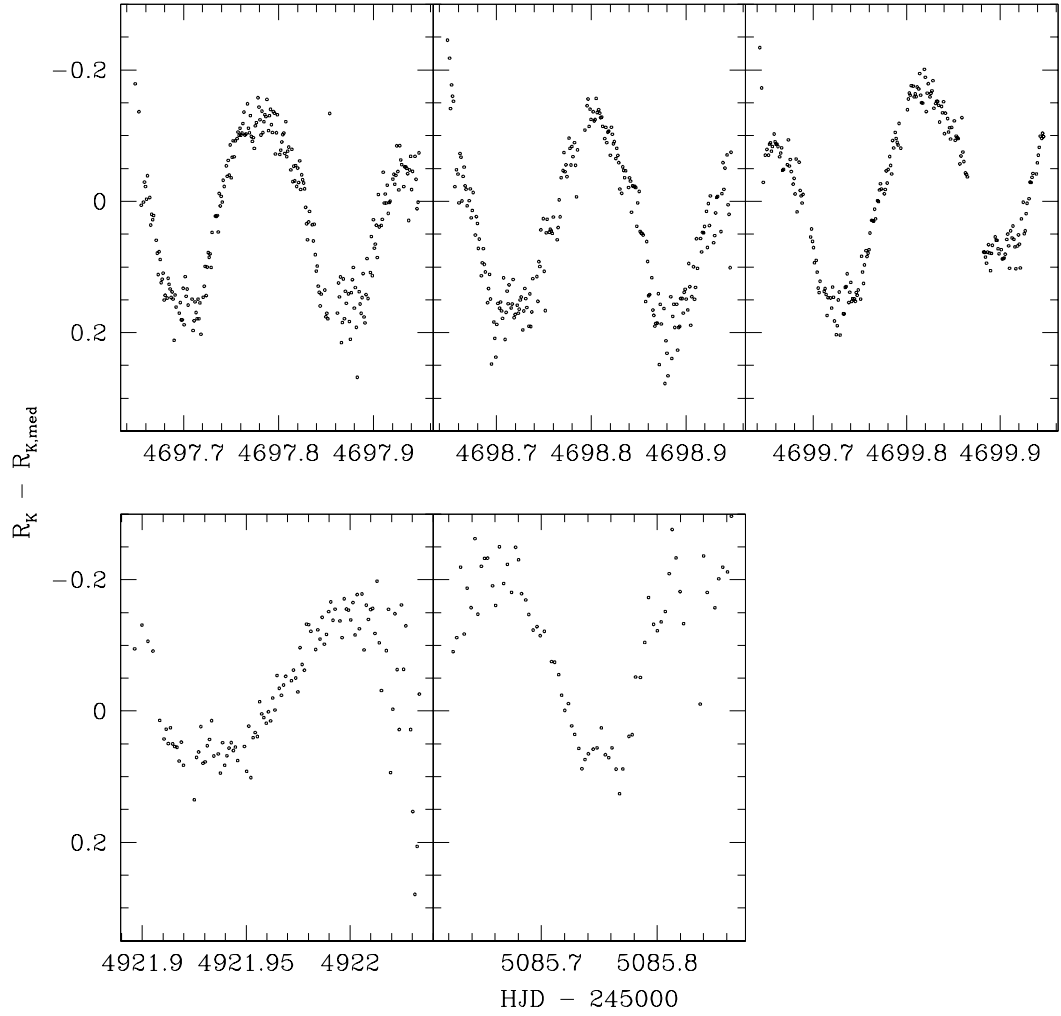}
\caption{Nightly light curves in $R_K$ for V2389 Cyg.\label{v2389}}
\end{figure}

{\bf V2389 Cyg:} Although this system does not appear to be a cluster member,
it does have total eclipses, as noted by \citet{street1}. We
note additionally that we detected a sudden change in the depth of one eclipse
from one night to the next (see Fig. \ref{v2389}). On the nights of HJD
2454697 through 2454699, we observed both eclipses on individual nights, and
while they were of equal depth on the first two nights, on the third night one
of the eclipses had increased in brightness by approximately 0.1 mag. We only
observed two later eclipses in the same filter band, but both were at the
brighter level.

\begin{figure}
\plotone{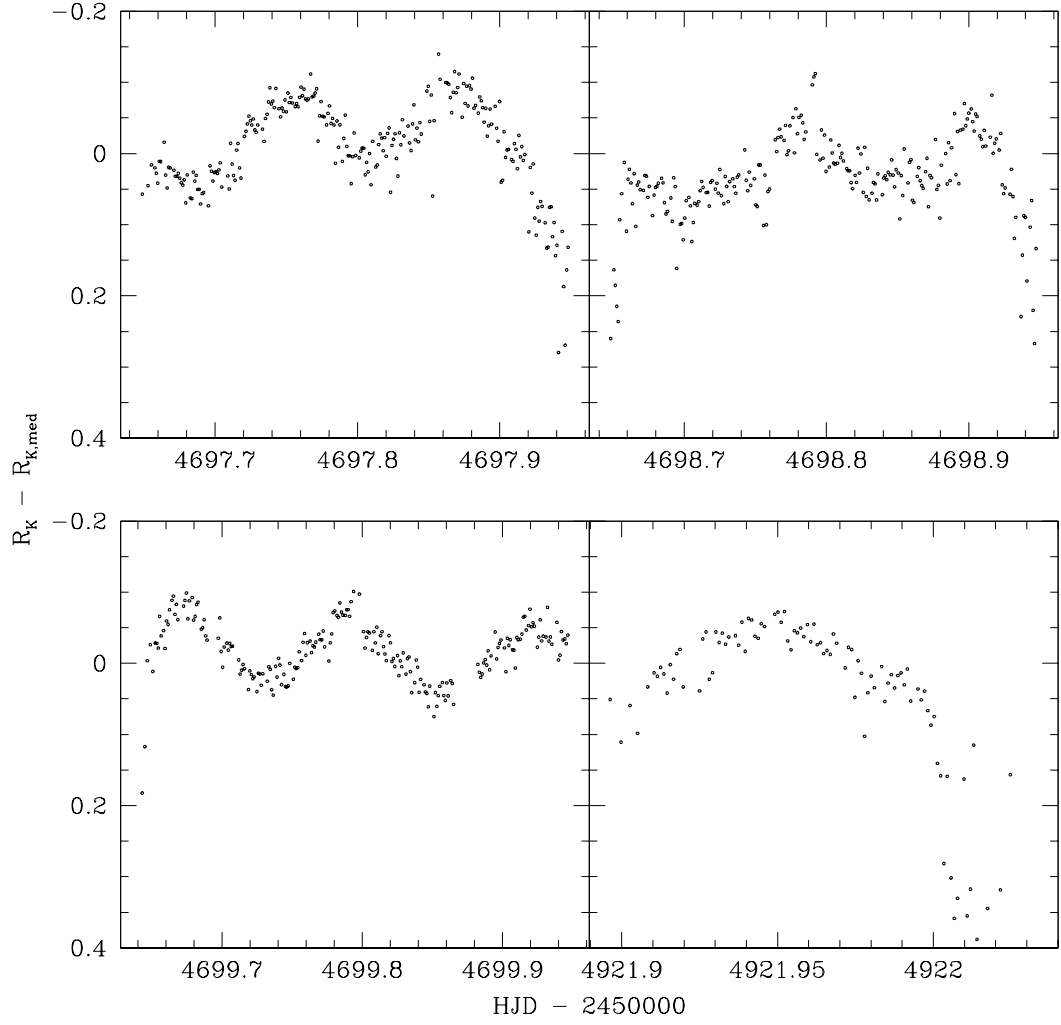}
\caption{Selected nightly light curves in $R_K$ for V2394 Cyg. \label{v2394}}
\end{figure}

{\bf V2394 Cyg:} According to the PLC analysis, this binary appears to be a
cluster member. The system shows short timescale ($\sim0.14$ d) variations
reminiscent of a contact binary, and long timescale variations in overall
brightness level (see Fig. \ref{v2394}). Both types of variations make it
difficult to derive an accurate period, but it is clear that it is close to
that of \citet{street1}.

\subsection{Detached Eclipsing Binaries (DEBs)}

Cluster members that are DEBs were one of the main goals of this study
because of their ability to provide masses and radii. Of the newly
detected DEBs, the most interesting are A494 (a possible blue
straggler), A259, A536, and A665 (near the cluster turnoff), and ID
1602 (faint main sequence). The stars below are ordered by brightness.

\begin{figure} 
\plotone{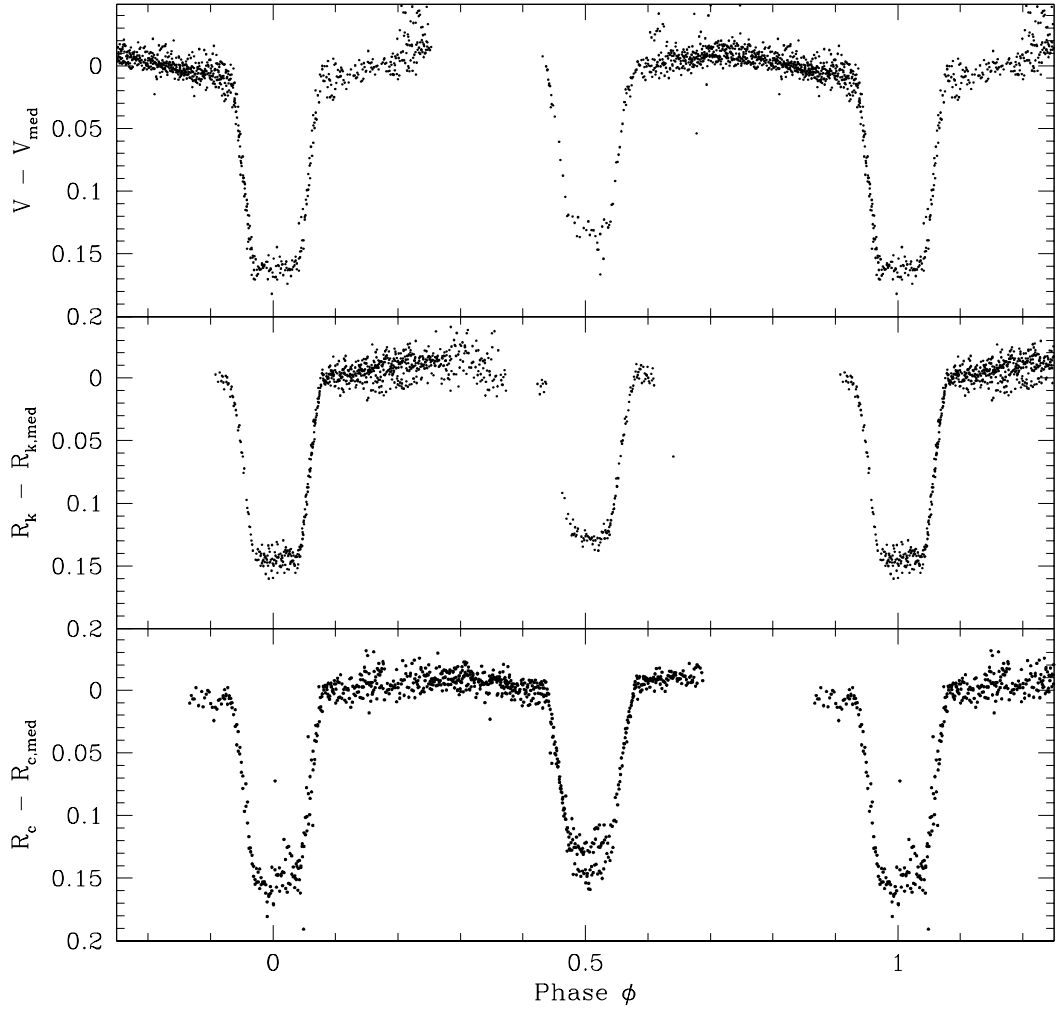}
\caption{Phased light curves for A494. \label{a494}}
\end{figure}

{\bf A494 (WOCS 007006):} Although this system was discovered to be variable
by \citet{ks}, they were unable to fully characterize the system due to their
small number of observations. For the first time, we detect clear evidence of
total eclipses in this short period binary, with only a small amount of
variation outside of eclipse (see Fig. \ref{a494}). There appears to be a
small difference in eclipse depths with this period, although we see some
variation in depth that may be an indication of spot activity. (Note that A494
was close to close to saturating pixels in the DAO images, so the variations
there should not be taken as necessarily representing the stellar
variability.) As a result, there is some possibility that the true period is
twice what we quote in Table \ref{vartab}.  The shallowness of the total
eclipses ($\sim 0.14$ mag) is a strong indication that there is a third star
present that contributes most of the system's light. We have two spectra taken
the HET (see Table \ref{rvtab}) that only show one clearly detectable
component. 

The system has identification number 59 in the proper motion study of
\citet{pm}, who gave it a 57\% probability of membership.  The radial
velocities of the brightest star at the two epochs are significantly different
from each other and one is significantly different from the cluster mean
($2.34\pm0.05$ km s$^{-1}$ with a dispersion $\sigma = 1.02\pm0.02$ km
s$^{-1}$; \citealt{hole}), indicating that it may be showing a reflex motion
in response to the gravity of the eclipsing binary. We can estimate that the
rotational velocity of the brightest star is $60 \pm 10$ km s$^{-1}$.  If it
is a cluster member, it is likely that the third component is a blue straggler
star in its own right, and the close binary contains relatively low-mass main
sequence stars. Using the eclipse depths as guides, we find that we can
reproduce the system photometry with two cluster main sequence stars with
$16.1 \la V \la 16.4$, which would in turn result in a third star ($V \approx
14.4$) falling near an extension of the cluster main sequence. That would make
A494 an example of a blue straggler system that must have formed from a
binary-binary star interaction. The straggler S1082 in the open cluster M67
presents a more extreme example in which 5 or more stars may have been
involved in producing the three stars that are observed today
\citep{vdb,s1082}.  Another example is the star 7782 in the older cluster NGC
188, which appears to involve two blue stragglers in a binary system
\citep{gell}. If the radial velocities of the stars in the eclipsing binary
could be measured, it might be possible to determine the masses of all three stars.



\begin{figure} 
\plotone{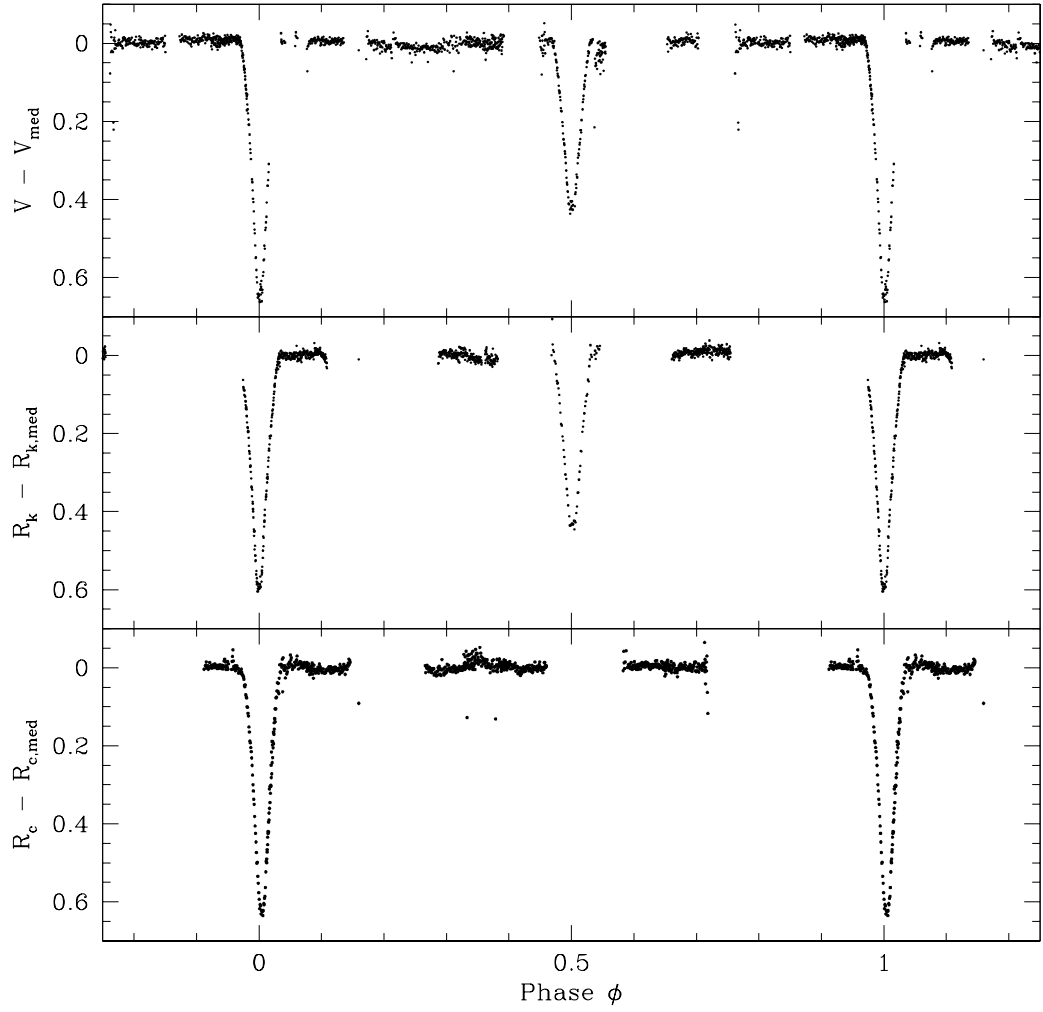}
\caption{Phased light curves for A259. \label{a259}}
\end{figure}

{\bf A259 (WOCS 040007):} There was an eclipse during some of our calibration
observations, although we had others on the same night that allowed us to
calculate reliable system values. Spectra show two components (A. Geller \&
R. Mathieu, private communication). The phased light curves are shown in
Fig. \ref{a259}. The observations in $R_K$ reveal that there is a short period
(about 45 min) of totality in both eclipses. Based on all of this information,
this system is an excellent candidate for deriving mass and radius information
for the component stars.

\begin{figure}
\plotone{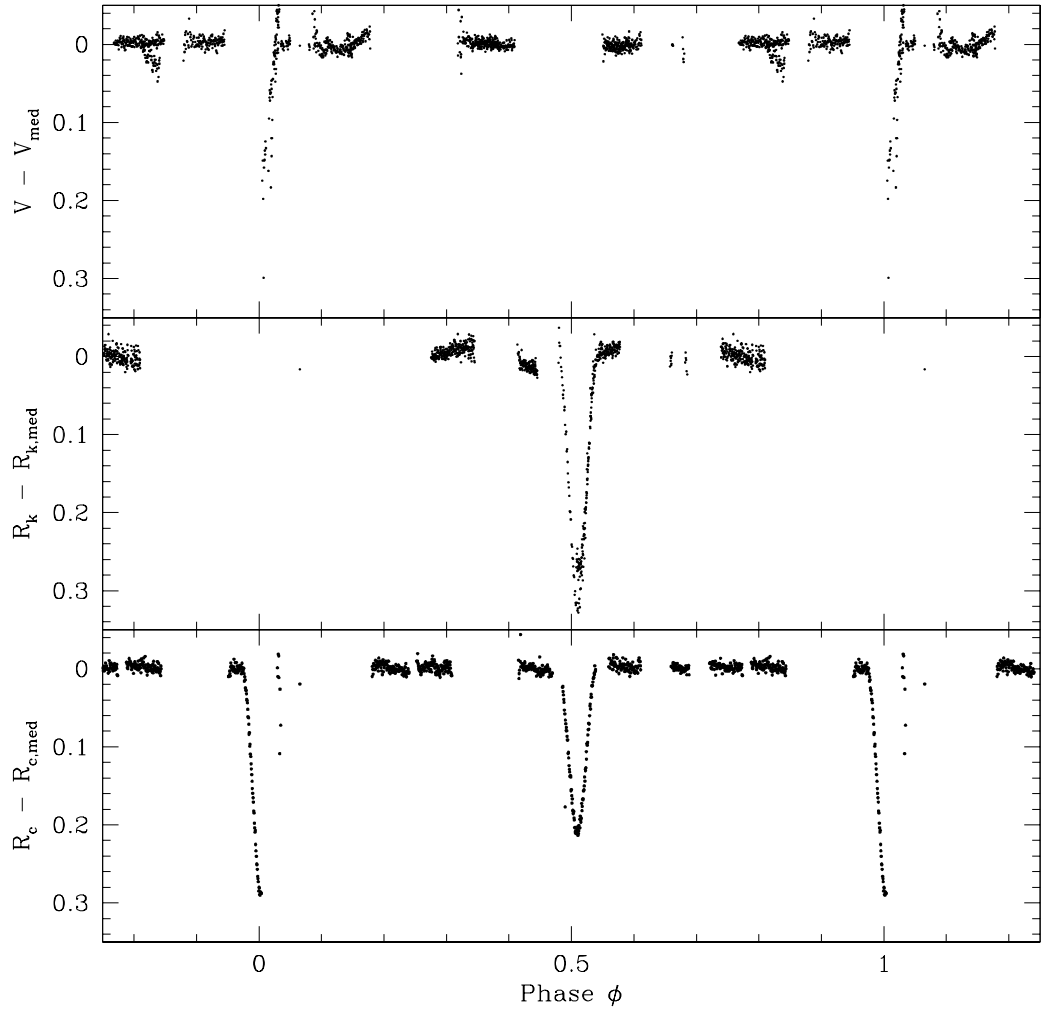}
\caption{Phased light curves for A536. The employed period comes from radial
  velocity observations (A. Geller \& R. Mathieu, private
  communication). \label{a536}}
\end{figure}

{\bf A536 (WOCS 022003):} The interval between two eclipses of different depth
observed in 2001 was about 2.12 d. However, the remainder of our observations
contain only partial coverage of several eclipses and were separated by
large time intervals.  Radial velocity information (A. Geller \& R. Mathieu,
private communication) allowed us to identify a period of 4.30103 d, which in
turn indicates that the eccentricity is near zero. (Phased light curves are
shown in Fig. \ref{a536}.) The radial velocities also indicate that the more
massive star is farther from us during the deeper eclipse in $R_C$, as
expected if both stars are on the main sequence.

A536 shows three components in spectra (A. Geller \& R. Mathieu, private
communication; Table \ref{rvtab}). Because the brighter stars are part of the
eclipsing binary, it should be possible to derive mass and radius information
for them. The third component may explain other peculiarities of the
photometric observations --- most notably that there are deep eclipses
observed in $R_K$ at the same phase when a clear secondary eclipse was
observed in $R_C$. We emphasize that radial velocities clearly rule out other
possible periods.  There is a star about $1\farcs6$ south of A536 that was not
resolved in $R_C$ that is probably influencing the measured eclipse
depths. This star appears to be WOCS 34003, which has $V = 15.66$ and $\bv =
0.62$.  Our images in $R_K$ have the best spatial resolution, which probably
results in less dilution of the binary's light.

The nearby star is also a possible contaminant in the spectra used for radial
velocities because in poor seeing conditions its light could enter the fiber
optic cables used in the spectrograph. (The cables have sky widths of
$2\arcsec$ for the HET spectra and $3\arcsec$ for the Geller \& Mathieu
spectra.) The third component in spectra appears near the cluster mean in
agreement with this idea. If true, this system could still provide a good age
constraint for the cluster. 

This object is an interesting test of how well image subtraction photometry
can do in deriving precise light curves in a moderately crowded stellar
environment. In a high-precision light curve from the {\it Kepler} mission,
these stars will undoubtedly be blended due to the nearly 4$\arcsec$
pixels. We are continuing work on this system with the aim of doing a full
analysis of the eclipsing binary. At worst it should be possible to treat the
nearby star as a ``third light'' contributor that dilutes the binary's light,
but with higher resolution observations the true eclipse depths should be
measurable.

\begin{figure}
\plotone{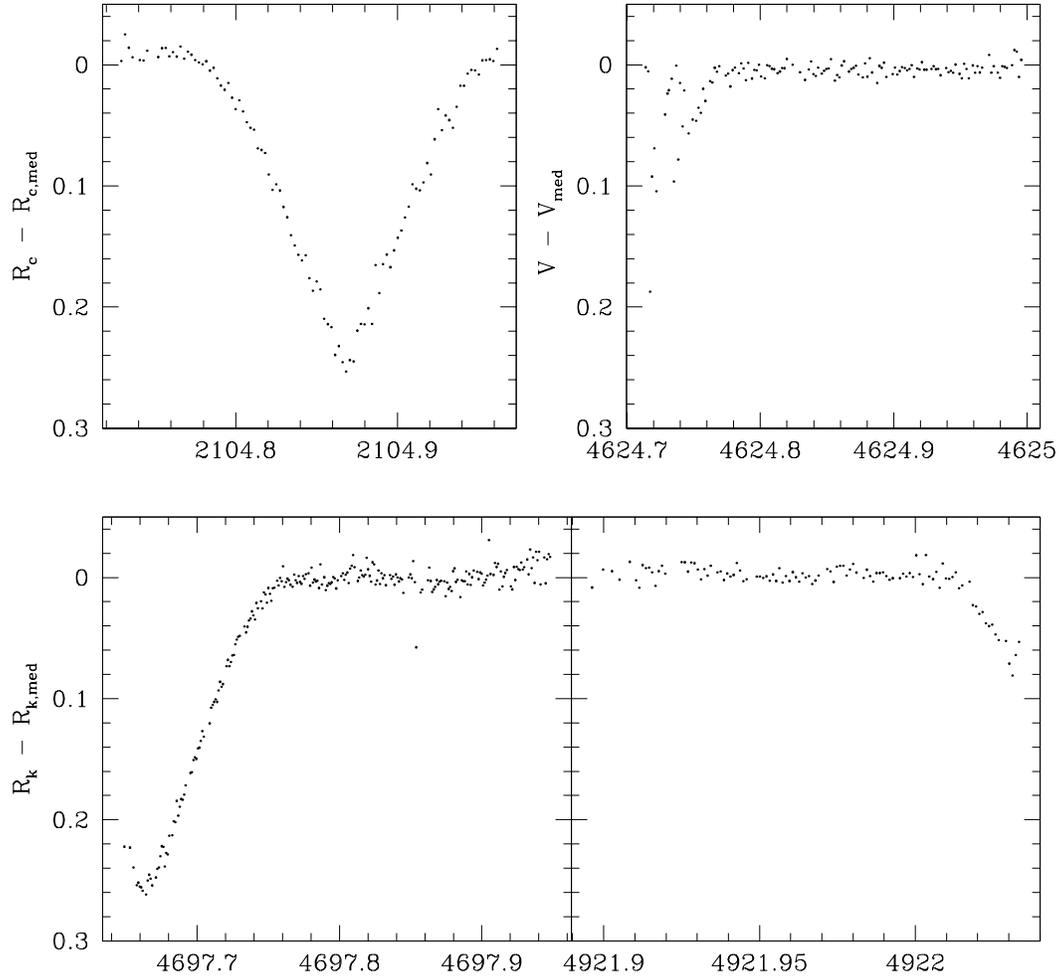}
\caption{Light curves for nights with detected eclipses for A665.\label{a665}}
\end{figure}

{\bf A665 (WOCS 024009):} We have not yet been able to identify a period for
the system due to a combination of partial coverage of eclipses and no
indication yet that the eclipses have different depths. Fig. \ref{a665} shows
our detections of parts of eclipses. High-precision {\it Kepler} light curves
should be able to identify differences in eclipse depths and enable a period
determination. In any case, this is likely to be the eclipsing binary with the
longest period detected in this survey. Spectra indicate that the system has
three components (A. Geller \& R. Mathieu, private communication). There is no
indication of a resolvable blended star, so this system may be a triple. At
least one of the brighter stars in the group is in the eclipsing binary, so it
should be possible to derive mass and radius information.

\begin{figure} 
\plotone{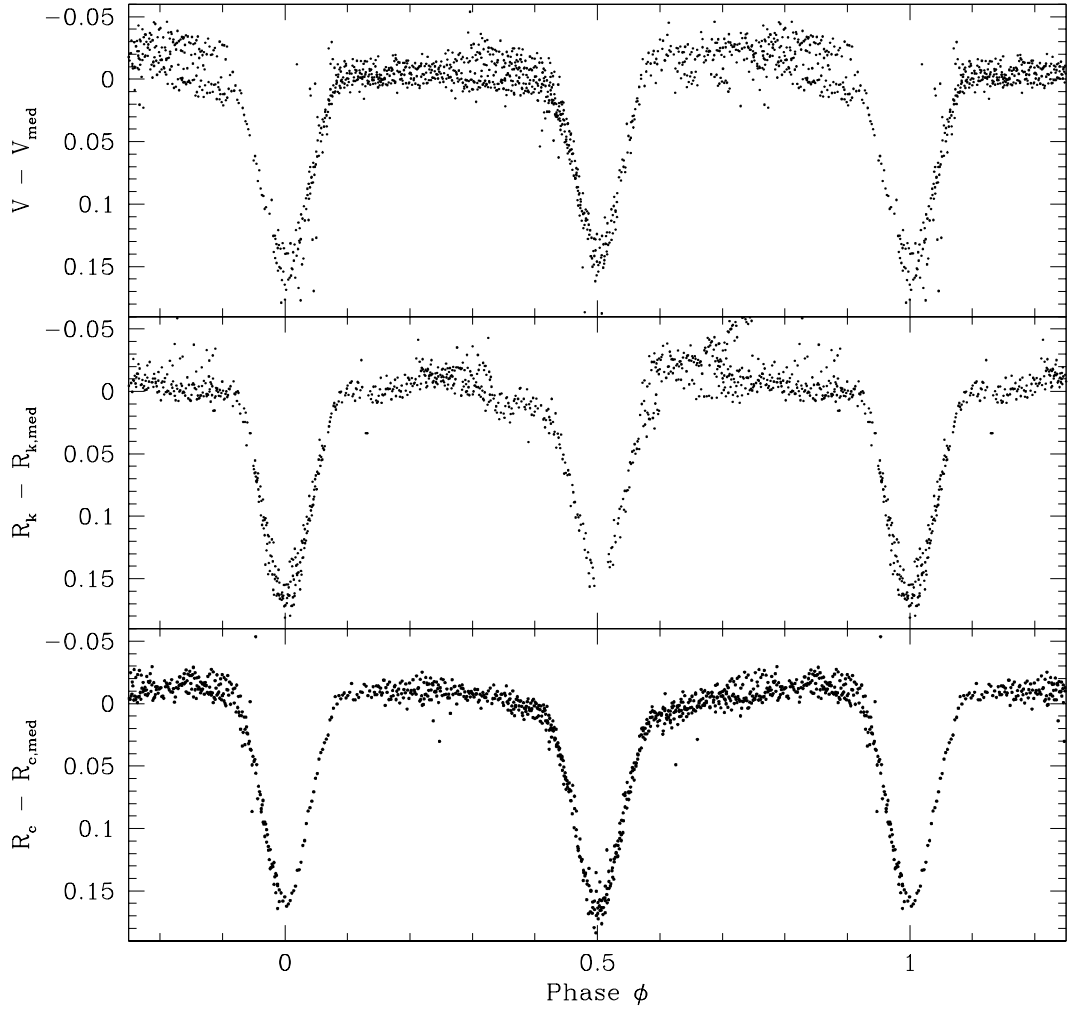}
\caption{Phased light curves for A824. \label{a824}}
\end{figure}

{\bf A824 (WOCS 059010):} This system shows eclipses of moderate ($\sim 0.15$
mag) depth (see Fig. \ref{a824}). Parts of eclipses were detected on 8 of the
10 nights of the DAO photometry observations, implying a period of a day or
less. The small size of the out-of-eclipse variation (as much as 0.02 mag)
indicates that the correct period is probably the longest possible acceptable
period. This led to the identified period of approximately 0.787 d. Because
the eclipses are of similar depth and because there are out-of-eclipse
variations (probably due to magnetic activity), it is hard to definitively
identify a primary and secondary eclipse.

In our best seeing images, this star appears to be a marginally
resolvable blend, with a separation of about $1\farcs2$ along an axis
ESE to WNW. The blend probably affects the depths of the eclipses in
all filter bands, but especially in the lower-resolution $R_C$ from
DAO. Spectra have only detected a cluster non-member with constant
radial velocity \citep{hole}. The detected eclipsing binary has a very
short period for a binary that shows little out-of-eclipse variation,
and so this reinforces the idea that the eclipsing binary is composed
of small faint main sequence stars. Because of the possibility the
binary could be a cluster member, we are studying this system further,
although the much brighter nonmember star will make spectroscopic
analysis more difficult.

{\bf V2381 Cyg:} A primary eclipse occurred on the night of our calibration
observations, affecting all of our measurements. We were able to correct the
$V$ measurements to out-of-eclipse values using the differential photometry,
but this is not yet possible with the $B$ and $I$ measurements. The quoted
$B-V$ color in Table \ref{vartab} comes from \citet{street1}, and puts the
system well to the blue of the main sequence. CMDs using instrumental
magnitudes from nights without eclipses confirm that the system is bluer than
the cluster main sequence. The implication is that the system is not a cluster
member.

\begin{figure}
\plotone{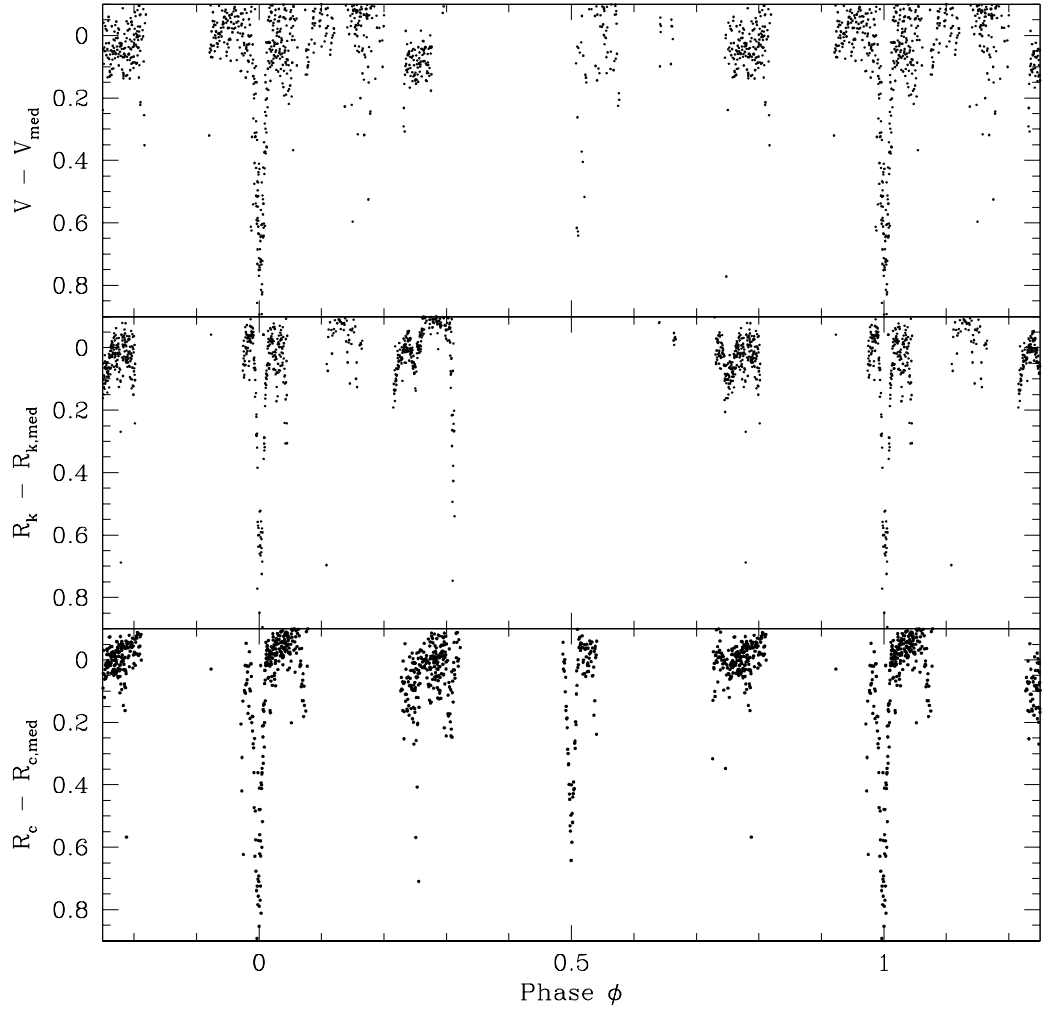}
\caption{Phased light curves for ID 1602.\label{1602}}
\end{figure}

{\bf ID 1602 (2MASS 19410280+4007116):} The photometry for this system places
it between the single-star main sequence and the equal-mass binary
sequence. We detected three deep ($\sim0.8$ mag) partial eclipses in $V$, one
shallower ($\sim0.6$ mag) eclipse in $R_K$, and part of four eclipses in
$R_C$. Three eclipses observed in $R_C$ were each separated by a little over
two days, and the middle one appears to be a shallower secondary eclipse. We
believe a period of 4.1108 d is most likely, and the data is phased to this
period in Fig. \ref{1602}.






\section{Conclusions}

One of the primary discoveries in this study was three variable
blue stragglers which each present evidence of formation
through binary-moderated dynamical interactions. These stars exist in
NGC 6819 alongside four additional blue straggler systems that show velocity
variability \citep{hole} indicative of binarity, although the history
of these systems cannot yet be reconstructed. These results emphasize
how photometric and spectroscopic studies of blue stragglers are
complementary --- photometric studies tend to be biased toward the
discovery of short-period binary systems, while spectroscopy is
essentially the only way of discovering populations of long-period
binaries \citep[e.g.]{gell}. However, together the photometric and
radial velocity variables only account for about 50\% of the blue
straggler population in this cluster. Spectroscopic abundance studies
\citep[e.g.]{shsa,ferr} provide another avenue for delving into blue
straggler formation histories even for single stars.

More importantly though, we have uncovered four detached eclipsing binary
systems near the cluster turnoff of NGC 6819. Three of the systems are very
likely to be cluster members, and the membership of the fourth is still to be
determined. Because of the selection bias against photometrically detecting
long period eclipsing binary systems, there is a fair chance that others are
present \citep[e.g.]{rkh}. The {\it Kepler} satellite provides an ideal means
of detecting additional systems in NGC 6819. If all of these systems are fully
analyzed to derive masses and radii for the eclipsing components, they will
densely constrain theoretical stellar models near the turnoff.

\acknowledgments We thank A. Geller and R. Mathieu
for valuable conversations on the spectra of some of the binaries, and
K. Brogaard for comments on the manuscript. We would like to thank the
Director of Mount Laguna Observatory (P. Etzel) for generous allocations of
observing time. Infrastructure support for the observatory was generously
provided by the National Science Foundation through the Program for Research
and Education using Small Telescopes (PREST) under grant AST 05-19686. We
would also like to thank A. Bostroem, C. Gabler, and J. B. Leep for assisting
with the photometric calibration observations.

This work has been funded through grant AST 09-08536 from the National Science
Foundation to E.L.S. A.T. gratefully acknowledges the support of the NSF
through grant AST 04-53609 as part of the CSUURE Research Experiences for
Undergraduates (REU) program at San Diego State University.

The Hobby-Eberly Telescope (HET) is a joint project
of the University of Texas at Austin, the Pennsylvania State University,
Stanford University, Ludwig-Maximilians-Universitat Munchen, and
Georg-August-Universitat Gottingen.  The HET is named in honor of its principal
benefactors, William P. Hobby and Robert E. Eberly.
This research made use of the SIMBAD database,
operated at CDS, Strasbourg, France, and the WEBDA database, operated at the
Institute for Astronomy of the University of Vienna.

\begin{deluxetable}{lcll}
\tablewidth{0pt}
\tablecaption{Photometry at Mount Laguna Observatory}
\tablehead{\colhead{Date} &
\colhead{Filters} & \colhead{mJD Start\tablenotemark{a}} & \colhead{$N$}}
\startdata
Mar. 4, 2001 & $V$ & 1973.994 & 53 \\
Jun. 5, 2008 & $V$ & 4623.715 & 137 \\
Jun. 6, 2008 & $V$ & 4624.714 & 134 \\
Jun. 7, 2008 & $V$ & 4625.715 & 96 \\
Jul. 21, 2008 & $V$ & 4669.659 & 168 \\
Aug. 18, 2008 & $R_K$ & 4697.649 & 246 \\
Aug. 19, 2008 & $R_K$ & 4698.649 & 231 \\
Aug. 20, 2008 & $R_K$ & 4699.645 & 218 \\
Oct. 6, 2008 & $VR_K$ & 4746.604 & 10,14 \\
Oct. 25, 2008 & $BVI_C$ & 4765.624 & 8,7,11 \\
Mar. 30, 2009 & $R_K$ & 4921.896 & 110 \\
Jun. 9, 2009 & $V$ & 4992.715 & 127 \\
Jun. 10, 2009 & $V$ & 4993.839 & 79 \\
Jun. 11, 2009 & $V$ & 4994.701 & 147 \\
Sep. 6, 2009 & $V$ & 5081.619 & 114 \\
Sep. 10, 2009 & $R_K$ & 5085.618 & 73 \\
Oct. 21, 2009 & $V$ & 5126.590 & 59 \\
Nov. 15, 2009 & $V$ & 5151.560 & 56 \\
\enddata
\label{phottab}
\tablenotetext{a}{mJD = HJD - 2450000.}  
\end{deluxetable}

\begin{deluxetable}{ccclllcccclll}
\rotate
\tabletypesize{\footnotesize}
\tablewidth{0pt}
\tablecaption{Variables Detected in the NGC 6819 Field}
\tablehead{\multicolumn{3}{c}{Identifications\tablenotemark{a}} & & & & & & & & & & \\
\colhead{WOCS/GCVS} & \colhead{A74} & \colhead{KS88/S02}&
\colhead{RA} & \colhead{DEC} &  \colhead{$P$ (d)} & \colhead{$V$} & \colhead{$\bv$} & \colhead{$V-I$} &
  \colhead{Type} & \colhead{$\Delta V$}& \colhead{Min. mJD\tablenotemark{b}} & 
  \colhead{Notes}}
\startdata
007002& 20pe &    &19:41:17.42&$+$40:11:03.6&0.155   &12.80&0.21&0.28&$\delta$ Scu&0.09&       &blue straggler?\\
007006& 494& V3 &19:41:03.29&$+$40:10:52.8&0.932209&14.08&0.28&0.38&EA&0.16,0.13&5081.805&total eclipses, blue straggler, poss. triple\\
013016&     &    &19:40:57.33&$+$40:04:55.7&0.666225&14.10&0.37&0.46&EB&0.55,0.22&5081.807&blue straggler\\
022003& 536 &    &19:41:12.07&$+$40:11:00.4&4.30103 &14.93&0.61&0.72&EA&0.18     &4698.655&phot. blend with nearby star\\
059010& 824 &    &19:41:06.43&$+$40:07:48.2&0.787121&15.10&0.72&0.81&EA&0.15     &4698.688&prob. blend with nearby star\\
024009& 665 &    &19:40:57.85&$+$40:09:27.5&        &15.17&0.65&0.79&EA&$\sim0.2$&2104.871&prob. triple\\
040007& 259 &    &19:41:33.94&$+$40:13:00.5&3.1851&15.65&0.71&0.88&EA&0.43     &4698.730&\\
052004& 355 & V1/V2&19:41:25.87&$+$40:12:22.2&0.348687&15.71&0.79&0.91&EW&0.19   &4697.727&\\
V2388 &  44 & 5834 &19:41:10.33&$+$40:15:18.3&0.366025&16.53&0.74&0.89&EW&0.33   &4697.705&\\
V2396 &     & 3856 &19:41:28.58&$+$40:16:24.8&0.293151&17.11&0.88&1.04&EW&0.22   &4697.746&\\
V2381 &     & 7916 &19:40:44.83&$+$40:09:23.0&1.468   &17.35&0.66&    &EA&1.3    &4994.832&prob. nonmember\\
V2393 &     & 4448 &19:41:22.61&$+$40:11:07.1&0.303209&17.37&0.97&1.09&EW&0.10   &4697.724&\\
V2389 &     & 5660 &19:41:11.73&$+$40:06:39.7&0.33847&18.04&0.58&0.68&EW&0.39    &4697.870&prob. nonmember\\
V2394 &     & 4441 &19:41:22.91&$+$40:14:39.5&0.2561 &18.19&1.01&1.37&EW&        &        &\\
1602  &     &      &19:41:02.80&$+$40:07:11.7&4.1108:&19.00&1.24&1.45&EA&0.8    &4994.74&\\
\enddata \tablenotetext{a}{WOCS: WIYN Open Cluster Survey; GCVS: General
  Catalogue of Variable Stars, \citet{gcvs}; A74: \citet{auner}; KS88:
  \citet{ks}; S02: \citet{street1}} \tablenotetext{b}{mJD = HJD - 2450000.}
\label{vartab}
\end{deluxetable}


\begin{deluxetable}{crrrrrrr}
\tablewidth{0pt}
\tablecaption{HET Radial Velocity Measurements}
\tablehead{\colhead{UT Date} &
\colhead{mJD} & \colhead{$v_A$ (\kms)} & \colhead{$\sigma_{A}$} & \colhead{$v_B$ (\kms)} & \colhead{$\sigma_B$} & \colhead{$v_C$ (\kms)} & \colhead{$\sigma_C$}}
\startdata
\multicolumn{6}{l}{A494:}\\
20091115 & 5150.56878 & $-39.5$ & 8.9 & 223.8: & 7.1 \\
20100602 & 5349.77955 & $-11.1$ & 6.3 &        &     \\
\hline
\multicolumn{6}{l}{A536:}\\
20091114 & 5149.56803 & $-64.3$ & 1.0 & 80.3 & 2.6 & $-3.4$ & 10.2: \\
\enddata
\label{rvtab}
\tablenotetext{a}{mJD = HJD - 2450000.}  
\end{deluxetable}

\clearpage

\begin{figure}
\plotone{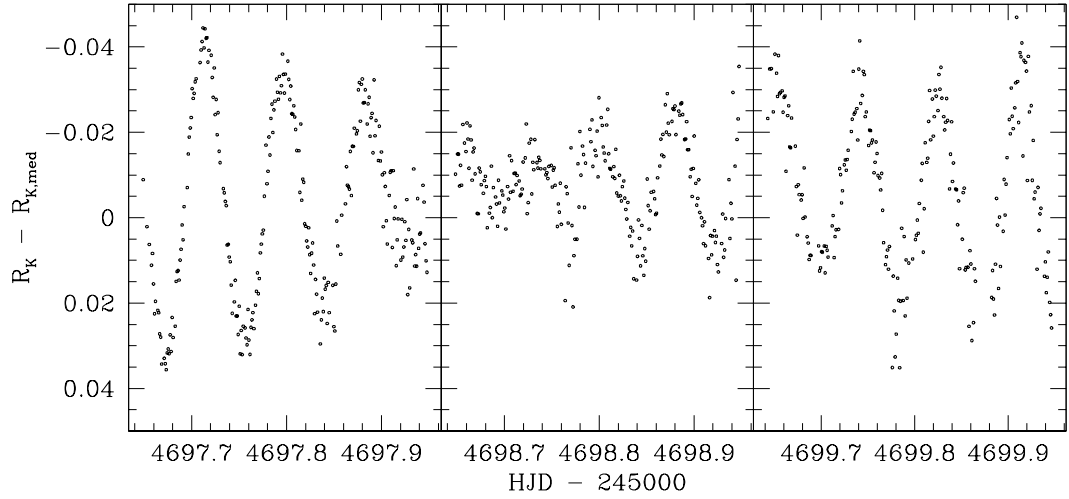}
\caption{Selected nightly light curves for the $\delta$ Scu variable A20pe. In
  this and all subsequent figures, the median for our observations in each
  filter band has been subtracted.\label{a20}}
\end{figure}

\end{document}